\documentclass{rrparticle}
\usepackage{graphicx}

\newcommand{\miktex}{\hbox{Mik\kern-.15em\TeX}}

\title{On the selection of models for the high-frequency quasiperiodic oscillations of black holes:\\ The mass problem} 
\author[1]{Ivan Zh. Stefanov}
\author[2]{Radostina P. Tasheva}
\affil[1]{Department of Applied Physics, Technical university of Sofia, \authorcr 8, Snt. Kliment Ohridski Blvd.,   1000 Sofia, Bulgaria, \authorcr Email:{ \em izhivkov@tu-sofia.bg}}
\affil[2]{Department of Applied Physics, Technical university of Sofia, \authorcr 8, Snt. Kliment Ohridski Blvd., 1000 Sofia, Bulgaria, \authorcr Email:{\em tasheva@tu-sofia.bg}}

\begin{document}
\maketitle
\begin{abstract}
High-frequency quasiperiodic oscillations of the X-ray flux comming from black-hole binaries have the potential to provide precise estimates of the mass and the spin of the central black hole, if an adequate model for them was available. There are several models in the literature but none of them is commonly accepted. One way to test the available models is to confront their predictions for the masses of black holes with results obtained through other methods. Here, we study the mass bounds that nine of the most commonly used models provide for three microquasars with known masses  GRS  1915+105,  GRO 1655-40 and XTE 1550-564 which display high-frequency quasiperiodic oscillations in their X-ray spectra. We also propose a statistical method for the assessment of the average success of the models. The results allow us to discard five of them. Here, "the mass problem" designates the conflict between their  predictions and the reference masses.

\end{abstract} 

\section{Introduction}
Quasiperiodic oscillations (QPOs) of the X-ray flux comming from black-hole binaries have been demonstrated to serve as a very precise method for the estimation of the mass and the spin of the central black hole. Recently, Motta et al. \cite{RP_to_GRO_Motta} applied the relativistic precession model to provide very precise estimates of the mass and the spin of the central object in the low-mass X-ray binary GRO 1655-40. An alternative study of the same object based on the same model was conducted in \cite{RP_GRO_Bambi}. It confirmed the results of the former work.

Which circumstances made this possible? The suitable choice of a model for the QPOs, the suitable choice of a metric of the spacetime, and the rare event of a simultaneous observation of C-type low-frequency QPOs and a pair of twin (a lower and an upper) high-frequency QPOs. The choice of a model and a metric, however, are far from unique. Even if we agreed not to question the Kerr metric, we would still face many uncertainties. One of the major difficulties with the application of the QPOs for the measurement of the basic parameters of black holes, mass and spin, is the absence of a verified and, hence, commonly accepted model for the high-frequency quasiperiodic oscillations (HF QPOs).

How could we test the different models for the HF QPOs and discard the improper ones? If sufficiently large observational data sets were available we could apply statistical tests to evaluate the models and reject the bad ones. Examples of such studies for the case of kHz QPOs, which are the analogues of HF QPOs for neutron stars, can be found in \cite{Lin2011_obs_confront_theory, mass_angular_relation_Chi2_2014, EoS_NS_QPO}. The situation with black holes is less favorable, since the HF QPOs observed in their power density spectra are very faint. As a results, HF QPOs are observed for a small number of black holes \cite{Klis_REVIEW, BHBs_McClintock, Xray_BHBs_Remillard, BHBs_Zhang, QPOs_universal_3:2_Zhou, QPOs_5000_BH}. Twin HF QPOs are even rarer. What is even worse, for most of these objects, unlike the situation with the kHz QPOs of neutron stars, only a single pair of twin HF QPOs is seen. So, it appears that no statistical testing is possible.

An alternative approach for the verifications of the models for the HF QPOs is to confront their predictions with facts that have been ascertained by other methods. Examples can be found in \cite{Stuchlik_confront} and \cite{SpinProblem}. These studies compare the constraints of the models for the HF QPOs on the spin of the black holes on the one side, to the values coming from other methods such as the fitting of the continuum of the X-ray spectrum and the fitting of the asymmetric profile of the ${\rm K}\alpha$ iron line, on the other side. A conflict between values obtained by the different methods has been ascertained. This conflict is termed ``the spin problem'' (See also \cite{NRM_RebuscoDifficulties}).

One of the caveats of the cited studies based on the comparison of predictions for the spin is the fact that they use the masses of the objects obtained through dynamical observations as input parameters. This is a possible source of uncertainty. Another potential problem is the fact that the different methods for the measurement of the  spin of the black holes might give conflicting results and one is not sure which value to take as a reference.  Differnt estimates for the spin of GRS 1915+105, for example, can be found in \cite{GRS_HF_QPOs_plus_0.98, GRS_Fe_line_spin, GRS_alternative_spin}.

In this work we propose a different approach. It is based on the comparison of the constraints that different methods impose on the mass, instead of the spin, of the black holes. What values do we use for the spin? An upper boundary on the spin comes from theory -- it cannot be greater than one. A lower bound on the spin can be obtained from the low-frequency quasiperiodic oscillations (LF QPOs) present in the spectra of the studied objects -- they originate in the accretion disk and are thus related to orbits which are outside of the innermost stable circular orbit (ISCO). To our knowledge, this condition has been applied for the first time in \cite{LF_radius_condition_Schnittman} to obtain lower bounds on the spin. It has been later applied in \cite{moyata} to test the models of the HF QPOs. Recently, \cite{Motta_LFQPOs_spin_constraints} used LF QPOs in a similar manner to constrain the spins of black holes.

Using the bounds on the spin we obtain bounds on the mass. The latter are confronted with the reference values obtained through dynamical observations. The idea that the confrontations of the predictions of the models for the HF QPOs for the masses of the observed black holes on the one side, with the dynamically obtained masses on the other side, are an indication of a bad model has been exploited by \cite{StuchlikKolosGRO}. They applied it to the microquasar GRO 1655-40. A possible outcome of the ascertained conflicts was proposed by the same authors in  \cite{StuchlikKolosGRO_explained}.

It appears that the mass constraints coming from some models for the HF QPOs fail to explain the observed masses. Can these failures be attributed to random effects? In attempt to find an answer to this question we propose a method for the measurement of the ``average success'' of the models. In brief, we normalize the predicted mass interval and the reference masses which allows us to make comparison and averaging between the different objects.

The experimental data for the object is given in the next section. The models are presented in Section~3. Then, follows a detailed description of the method that we use to obtain conservative lower bounds on the spins of the black holes. The bounds on the masses of the studied objects that we obtain are in Section~5. Section~6 contains the assessment of the quality of the models based on their average success. Summary of the results can be found in the last section.  The main formulas are in Appendices A and B. 
\section{Objects}\label{objects}
The test that we propose requires black holes with known masses the power density spectra of which have the following components: $C$-type LF QPOs\footnote{We refer the reader to \cite{ABC} and \cite{Motta_REVIEW} for the classification of LF QPOs.} and a pair of twin HF QPOs in 3:2 ratio. To our knowledge,  there are only three microquasars which meet this requirements: GRS 1915+105, GRO 1655-40 and  XTE 1550-564. The masses of their black holes, the frequencies of the LF and HF QPOs and references for them are commented on in this section and summarized in Table~\ref{objects}.
\begin{table*}
	\begin{minipage}{140mm}
		\caption{Data for the studied black holes}
		\small
		\begin{tabular}{|c|c|c|c|c|c|c|c|}
			\hline
			source          & $M/M\odot$ & Ref.  &  $\nu_{ \rm LF}$, [Hz] & Ref. & $\nu_{ \rm L}$, [Hz]  &$\nu_{ \rm U}$, [Hz]& Ref.\\
			\hline
			&&&&&&&\\
			GRS 1915+105    & $12.4_{-1.8}^{+2.0}$ & \cite{GRS_12_4_mass} &  $8.107\pm 0.043$ & \cite{GRS_LFQPOs_2016_C_state_QPOs} & $113 \pm 5$    &$168 \pm 3$  & \cite{GRS_HF_QPOs_plus_0.98} \\[2ex]
			\hline
			&&&&&&&\\
			XTE 1550-564    &  $9.1 \pm 0.6$ & \cite{XTE_mass} &  $18.037_{-0.067}^{+0.069}$ & \cite{RP_to_XTE_Motta} & $184 \pm 5$    &$276 \pm 3$  & \cite{XTE_HF}\\[2ex]
			\hline
			&&&&&&&\\
			GRO 1655-40     & $5.4 \pm 0.3$$^{a}$ & \cite{GRO_Beer_54} &  $28.3_{-0.1}^{+0.1}$  & \cite{RP_to_GRO_Motta} & $298\pm4$    &$441\pm2$  &  \cite{RP_to_GRO_Motta} \\[2ex]
			\hline
		\end{tabular}\label{objects}
	\end{minipage}
	\begin{minipage}{140mm}
		${}^a$ \small{ An alternative estimate of the mass of GRO 1655-40, $6.30 \pm 0.27 M_\odot $, is given in \cite{GRO_6_3_mass, GRO_spin_continuum_plus_mass}}
	\end{minipage}
\end{table*}
\subsection{GRS 1915+105}
Twin HF QPOs the ratio of which is close to 3:2 in the spectrum of GRS 1915+105 have been reported in \cite{GRS_113_165_BAAS}. The values that they report are 165~Hz and 113~Hz. The presence of this pair according to the authors of \cite{GRS_113_165_BAAS}  suggests that it is produced by the same mechanism that produces the 3:2 HF QPOs of GRO J1655-40 and XTE J1550-564. They also say that the 3:2 pair provides additional support for the conjecture that the fast X-ray variability of black hole binaries is a relativistic effect and is related to some kind of resonance. The values that we choose to work with are 113~Hz and 168~Hz which were later cited in several papers coauthored by Ronald Remillard including \cite{GRS_HFQPOs_Remillard2004}, \cite{GRS_HF_QPOs_plus_0.98} and in a chapter of a monograph devoted to compact stellar X-ray sources \cite{BHBs_McClintock}.

The HF spectrum of GRS 1915+105 was later reanalyzed in \cite{GRS_Belloni2006_no_113} and \cite{GRS_Belloni2013_no_113}. We should note that these studies found no evidence of the 113~Hz QPO.

GRS 1915+105 is a peculiar source. Beside the two HF QPOs cited above, its HF spectrum contains several more: $\sim 27$, $\sim34$, $\sim 41$ and $\sim67$. The last one appears to be the most persisted and the most clearly seen. As noted by \cite{GRS_Belloni2006_no_113} the values $27:41:67$ are in $2:3:5$ ratio which is once again reminiscent of resonance.  In order to explain the full set of HF QPOs we need a more elaborate model such as the one proposed in \cite{GRS_Stuchlik_more_elaborate_model}, for example. Here our considerations  are constrained to the 3:2 pair since the authors of the models that we want to test have aimed at the explanation of exactly this pair.

A recent study of the LF QPOs from all the RXTE observations of GRS 1915+105 is \cite{GRS_LFQPOs_2013}. The highest value that can be found in Table 1 in this paper is 7.987~Hz. The observation has been made on November 15, 1999 or MJD 51497. The ABC classification of the LFQPOs is not given. The authors of \cite{GRS_LFQPOs_2013}, however, cite an earlier paper by \cite{GRS_LFQPOs_ABC_classification} according to which on that day the X-ray source is partially in $C$ state. The same QPO but with a slightly different value of the centroid frequency 7.943~Hz has been reported also in \cite{GRS_LFQPOs_2016_C_state_QPOs}. According to the authors of this paper the QPOs produced when the X-ray source is in C state are probably equivalent to $C$-type LF QPOs. This paper gives another, slightly higher value: 8.107~Hz. The observation was made on July 7, 1996 or MJD 50271. We choose to work with this value.

The most recent estimate of the mass of the source that we could find is made by \cite{GRS_12_4_mass}. According to this work the mass of GRS 1915+105 is $12.4_{-1.8}^{+2.0} M\odot$.
\subsection{XTE 1550-564}
A pair of HF QPOs in the power-density spectrum of XTE 1550-564 with frequencies in nearly 3:2 ratio was reported by \cite{XTE_HF} -- a upper one at $\sim270$ Hz and a lower one at $\sim180$ Hz. These HF QPOs have been reexamined by \cite{XTE_HF}. The second HF QPO $\sim 180$ Hz is significant (5.0 $\sigma$) after averaging of 12 observations. We will work with the values that they provide, reported here in Table~1.  

A much later study \cite{Belloni_XTE_no} found no confirmation for the presence of simultaneous HF QPOs in the power density spectrum of this object. In this work, however, the different observations have been analyzed separately, without any averaging, which, as the authors state, reduces the sensitivity of the detection. 

For our study simultaneity is not obligatory. It is important as long as we treat it as an evidence for the presence of two different HF QPOs in the power density spectrum of a given object. Furthermore, we hypothesize that the simultaneous pair is born on the same orbit in the accretion disk. If this was not true our method for the obtaining of lower bounds in the spin of the black holes,  presented in Section 4, would be invalid. 

Even though the presence of a simultaneous pair of HF QPOs in the power density spectrum of XTE 1550-564 is questionable, this object still fits in our scheme as long as it has two different HF QPOs  which we assume to originate on the same obit and the frequencies of which are described by one of the models listed in Table~2.

We should also note, however, that even the presence of the two different HF QPOs is not certain for this object as Méndez et al. show in \cite{XTE_not_simultaneous}. The  $\sim~270$ Hz and the $\sim~180$ Hz QPOs, according to this study,  are the same QPO seen at different frequencies in different observations. Their study, however, did not take into account the entire set of observations in which the previous studies report two differnt QPOs. 
\subsection{GRO 1655-40}
GRO 1655-40 is the only black-hole binary (BHB) whose power density spectrum contains an authentic simultaneous 3:2 pair of HF QPOs. In all other sources the 3:2 pair is discernible only after averaging of the data from several observations. The most recent values of the two simultaneous HFQPOs can be found in \cite{RP_to_GRO_Motta}: $298\pm4 \,{\rm Hz}$ and $441\pm2 \, {\rm Hz}$. The same paper also gives an exhaustive list of the $C$-type LF QPOs of GRO 1655-40. Here we need just the highest one: 28.3~Hz.

For the mass of GRO 1655-40 the authors of \cite{RP_to_GRO_Motta} cite the value $5.4\pm 0.3 M\odot$ which had been obtained by \cite{GRO_Beer_54} through spectro-photometric optical observations. One should note, however, that there is an alternative and more recent estimate of the mass of this black hole, $6.30 \pm 0.27 M_\odot $,  given in \cite{GRO_6_3_mass}. We will work with the former value. Bellow we add a comment on the effect that an alternative reference mass which differs from the value that we have chosen by $\sim 1 M\odot$ would have on our results and conclusions.

\section{Models for the HF QPOs}
Most hypotheses concerning the origin of QPOs assume a direct connection between  the QPO frequencies and the fundamental frequencies of motion of matter orbiting black holes (BH). The models which consider either ``hot spot'' or ``disk oscillation'' modes comprise two large groups  -- relativistic precession (RP) and resonant models.
The frequencies that the different models attribute to the observed lower and upper HF QPOs are summarized in Table \ref{table_models}.
\begin{table}
	\caption{Models for the HF QPOs.}
	\begin{center}
		\begin{tabular}{ |c|c|c|}
			\hline
			Model &$\nu_{\rm L}$& $\nu_{\rm U}$ \\
			\hline
			~~3:2 & $\nu_{\rm r}$ & $\nu_{\rm \theta}$ \\
			~~3:1  & $\nu_{\rm \theta}-\nu_{\rm r}$ & $\nu_{\rm \theta}$ \\
			~~2:1   & $\nu_{\rm \theta}$ & $\nu_{\rm \theta}+\nu_{\rm r}$ \\
			~~3:2 K  & $\nu_{\rm r}$ & $\nu_{\rm \phi}$  \\
			~~ RP, 3:1~K  & $\nu_{\rm \phi}-\nu_{\rm r}$ & $\nu_{\rm \phi}$\\
			~~TD, 2:1~K  & $\nu_{\rm \phi}$ & $\nu_{\rm \phi}+\nu_{\rm r}$ \\
			~~RP1   & $\nu_{\rm \phi}-\nu_{\rm r}$ & $\nu_{\rm \theta}$ \\
			~~RP2   & $\nu_{\rm \phi}-\nu_{\rm r}$ & $2\nu_{\rm \phi}-\nu_{\rm \theta}$  \\
			~~WD    & $2(\nu_{\rm \phi}-\nu_{\rm r})$ & $2\nu_{\rm \phi}-\nu_{\rm r}$ \\
			\hline
		\end{tabular}
	\end{center}
	\label{table_models}
\end{table}
\subsection{Relativistic precession model (RP)}
The precession models predict QPOs excitation from a variety of resonances between the precession and the orbital frequencies under certain conditions, for example inhomogeneities orbiting close to the inner disk boundary.

The relativistic precession model proposed by \cite {RP_StellaVietriModel}, \cite{RP_StellaVietriMorsink} and \cite{RP_StellaVietriKerr}  explains the twin HF QPOs as a result of relativistic epicyclic motion of radiating hot blobs traveling along orbits with different radii $r$ in the inner part of the accretion disk.
Their correlation can be attributed to the periastron precession of the relativistic orbits occurring in the strong gravity field in the vicinity of BH. This model associates the upper of the twin HF QPOs with the orbital frequency, $\nu_{ \rm U }=\nu_{ \rm \phi}$\footnote{See appendix \ref{app_freqs} for the explicit form of the frequencies met in this section.}. The lower one is attributed to the periastron precession frequency, $\nu_{ \rm L}=\nu_{ \rm per}=\nu_{ \rm \phi}-\nu_{ \rm r}$ . To match the observed 3:2 ratio the HF QPOs have to be generated very close to the ISCO. According to authors of the relativistic precession model the LF QPOs are a result of Lense-Thirring precession\footnote{The idea that C-type LF QPOs have geometric origin is supported by the original papers of the authors of the RP model \cite{RP_StellaVietriModel}, \cite{RP_StellaVietriMorsink}. There is a number of studies in the context of neutron stars that question this interpretation (See, for example, \cite{Not_Frame_Dragging}). A recent example is the thorough research conducted by \cite{Klis_new_LFQPOs}. Instead of one, they found several frequencies in the low-frequency sector that could be identified as LF QPOs. Frame dragging, i.e. Lense-Thirring precession, could explain the presence of only one of them.} with nodal frequency $\nu_{ \rm  nod}=|\nu_\phi-\nu_\theta|$. On the dependence between LF and HF QPOs we refer the reader also to \cite{PBK}. The relativistic precession model is successful in explanation of this relation for a number of neutron stars and black holes \cite{Klis_REVIEW}. These models, however, do not provide generic explanations for the observed 3:2 frequency ratio R=3/2 of the twin HF QPOs. Hence, it is a good idea to explore, if some orbits are preferred to others.

\subsection{Tidal disruption model (TD)}
Another  ``hot spot'' model is the tidal disruption model (TD), where hot orbiting clumps distorted by the tidal forces of the black hole and forming ``ring-like'' segments are responsible for the observed modulation of the power density spectra \cite{TD1, TD2,  TD3}. The TD model implies that the frequency ratio R is in the range $(1,2)$ for any spin of the black hole and radius of the orbit but still does not firmly constrain the ratio of the twin HF QPO \cite{Stuchlik_confront}. The Roche radius $ r_{ \rm TD}$ is the distance where tidal forces start to disrupt an approaching object. Its upper limit for a fluid body with density $\rho$ orbiting a massive body with mass $M$ is $ \propto (M/\rho)^{ \rm 1/3}$. The corresponding  frequency according to Kepler's third law is $\nu_{ \rm TD} \propto(GM/r_{ \rm TD})^{ \rm 1/2}$. After $ r_{ \rm TD}$  is replaced $\nu_{ \rm TD}$ is proportional to $(G\rho)^{ \rm 1/2}$.
As a result $\nu_{ \rm L} =\nu_{ \rm TD}\propto10^{ \rm -3}$Hz, which is much lower then the observed ~100 Hz frequency. Nevertheless, the  authors of this model managed to obtain plausible light curves and to fit the HF part of power density spectra of the low mass X-ray binary  XTE 1550-564. Moreover, with this model they were able to mimic the twin HF QPOs.

\subsection{Warped disk model (WD)}
Kato introduced for the first time the warped disk resonance model which proposes an excitation mechanism for the orbiting particle oscillations \cite{Kato2004a, Kato2004b, Kato2005a, Kato2005b, Kato2007}. The internal viscid and adiabatic perturbations of a deformed accretion disk lead to a horizontal resonance responsible for HF QPOs. The model is successfully applied to estimate the mass and angular momentum of LMXB GRS 1915+105 \cite{Kato2004c} .

\subsection{RP1 and RP2 - improved precession models}
These models are versions the precession models, i.e. the QPOs are supposedly excited by resonances of precession and nonaxisymetric oscillation modes. They are connected to  oscillation modes whose frequencies attributed to slow rotation coincide with the frequencies predicted by RP model. In the case of RP1 the rotating fluid torus has slight eccentricity and its vertical oscillations are superimposed on the precession modes to finally form the emitted radiation flow \cite{RP1}. The author expresses the lower frequency as the relativistic periastron precession frequency  $\nu_{ \rm L}=\nu_{ \rm per}=\nu_{ \rm \phi} -\nu_{ \rm r}$, while the upper one is hypothesized as $\nu_{ \rm U}=\nu_{ \rm \theta}$. The RP2 model according to \cite{Stuchlik_confront} and \cite{mass_angular_relation_Chi2_2014} is based on perturbations, where $\nu_{ \rm \phi}$  is  not the dominant frequency.  HF QPOs are produced by the resonance between radial and vertical modes. The RP1 model can explain the twin HF QPOs in XTE 1550-564 \cite{Stuchlik_confront}. It also gives reasonable explanation of the QPOs in GRO J1665-40 and provides results similar to \cite{RP_to_GRO_Motta}.

\subsection{Nonlinear resonance models (NRM)}
There are two types of resonant disk oscillation models -- the epicyclic resonance and the warped disk oscillation (WD) models. Resonant epicyclic models of the twin HF QPOs \cite{NRM_Abramowicz, NRM_AbramowiczSpinEstimate, NRM_AbramowiczTheory} suggest resonance between the fundamental oscillation modes of the accretion disk. The assumed resonance is either parametric, or a nonlinearly forced one. The model predicts that one of the frequencies $\nu_{ \rm -}= \nu_{ \rm \theta} -\nu_{ \rm r}$  and $\nu_{ \rm +}= \nu_{ \rm \theta} +\nu_{ \rm r}$  is in 3:2 ratio with the vertical frequency $\nu_{ \rm \theta}$.  For $\nu_{ \rm \theta}:\nu_{ \rm r} = 2:1$, $ \nu_{ \rm L}=\nu_{ \rm \theta}$, $\nu_{ \rm U}= \nu_{ \rm +}= \nu_{ \rm \theta}+\nu_{ \rm r}$. For $\nu_{ \rm \theta}:\nu_{ \rm r}= 3:1$,  $\nu_{ \rm L}=\nu_{ \rm -} =\nu_{ \rm \theta} -\nu_{ \rm r}$, $\nu_{ \rm U}=\nu_{ \rm \theta}$.
The commensurability of the frequencies is crucial for the NRM models. Although attractive for their simplicity these models cannot provide an adequate mechanism for the excitation of the QPOs\footnote{We should mention for the reader, however, that excitation mechanisms (more than one) have been proposed in \cite{NRM_excitation_mechanism}.}. There is also a discrepancy \cite{SpinProblem, NRM_RebuscoDifficulties} between their predictions for the angular momenta and measurements based on other methods such as: jet emission analysis, analysis of the profile of the $K_{ \rm \alpha}$  iron line and spectral continuum fitting. We refer the reader to \cite{AlievKerr} and \cite{NRM_Bambi_NRM_non_Kerr} for applications of the NRM to Kerr and non-Kerr black holes, respectively. Nonlinear resonances occurring in the field of braneworld Kerr black holes and Kerr superspinars (i.e. Kerr naked singularities) have been studied in \cite{NRM_Braneworld} and \cite{NRM_superspinars}.

\subsection{Keplerian nonlinear resonance model}
The Keplerian nonlinear resonance model differs from the epicyclic NRM in its prescription for the  upper frequency of the HF QPOs. In the former model $\nu_{ \rm U} = \nu_{ \rm \phi}$  instead of $\nu_{ \rm U} = \nu_{ \rm \theta}$.

\section{Method for the Attaining of the Mass Bounds}\label{Method}
The method that we apply to obtain loose bounds on the mass of the central object in black-hole binaries has been proposed first in \cite{moyata} but is presented in the current section in a different way. It has been designed for the Kerr metric (and takes advantage of the qualitative behavior of the fundamental frequencies of test particles and other functions derived from them) but could be generalized also to other black-hole spacetimes which are parameterized by two parameters. In the case of Kerr spacetime the two parameters are mass $M$ and specific angular momentum (or spin) $a$.

The main idea of the method is to pose constraints on the spin of the black hole and use the mass-spin relation resulting from a model for the HF QPOs to constrain the mass.

The current study is based on the following assumptions:
\begin{enumerate}
	\item \textbf{Kerr black hole}
	
	The central object of the microquasars is a Kerr black hole. Its specific angular momentum takes values in the interval $a\in[0,1]$.
	
	\item \textbf{LF QPOs due to Lense-Thirring precession}
	
	The (C-type) LFQPOs occur due to Lense-Thirring precession and $\nu_{\rm LF}=\nu_{\rm LT}$. According to the RP model LF QPOs result from the Lense-Thirring precession of matter inhomogeneities in the accretion disk, dubbed hot spots and treated as point particles. The variance of the frequency values is produced by the variance of the orbital radii of the hot spots. The frequency of the LF QPOs is associated with the nodal precession frequency $\nu_{ \rm  nod}=|\nu_\phi-\nu_\theta|$.
	
	Alternative models of the LF QPOs are proposed by \cite{PrecessingRing}, \cite{Ingram_Done_model} and \cite{Ingram_unified_model}. They associate the LF QPOs with the precession of thick disc which is formed by the hot inner accretion flow and which rotates as a solid body. The authors of \cite{PrecessingRing} assumed that the thick disc orbits the black hole along geodesic trajectories. In other words the precession frequency of the thick disc is given by the LT frequency of a single free particle. In the model developed in \cite{Ingram_Done_model} and \cite{Ingram_unified_model} the precession frequency of the thick disc is given by the weighted average of the LT frequencies of all test-particle circular orbits in the region that it spans.
	
	The test particle precession model and the global rigid precession model have been compared recently in \cite{Motta_global_precession}.

	\item \textbf{prograde motion}
	
	Retrogrde motion is highly unlikely. Arguments can be found in \cite{disfavour_retrogade}. This fact has been noted also by \cite{Motta_LFQPOs_spin_constraints}.

	\item \textbf{a pair of twin HF QPOs in 3:2 ratio}
	
	The authors of the nonlinear resonance models and the Keplerian nonlinear resonance models aimed at the explanation of the 3:2 pair of HF QPOs which they assumed to be an experimentally ascertained fact. The author of the Warped disk model also stresses on the importance of the commensurability of the HF QPOs and interprets it as an indication of some kind of resonance.
	
	The 3:2 ratio is not an integrable part of all of the models that we consider here, however. The RP model, and its modifications RP1 and RP2 are more general. They are not bound the 3:2 ratio. They can be applied to pairs of HF QPOs which are not in 3:2 ratio. In cases of microquasars the spectra of which contain more than two HF QPOs such as GRS 1915+105, however, it is unclear which pair of the observed frequencies to be identified with those predicted by the RP model or its modified versions. The situation gets even more complicated, if the observed HF QPOs are not simultaneous. We choose to apply the RP model and its modified versions to the pair of HF QPOs the ratio of which is close to 3:2. In the past the RP models have been applied, for example, to model the 3:2 pair of HF QPOs of GRO~1655-40 in \cite{StuchlikKolosGRO} and \cite{StuchlikKolosGRO_explained} and of GRS  1915+105 in \cite{Stuchlik_confront}. One more reason to include the RP model in our list, even though it is not aimed at and constrained to the 3:2 pair, is that the prescription that it makes for the frequencies of the HF QPOs coincides with that of the 3:1 K model which, as mentioned above, has been designed to explain exactly that pair of HF QPOs.
	\item \textbf{QPOs are born in the accretion disk}
	
	LFQPOs originate at (unknown) orbits outside of the innermost stable circular orbit (ISCO) on orbits with radii greater than $r_{\rm ISCO}$, the radius of the ISCO.
	
	\item \textbf{not necessarily on the same orbit}
	
	We do not require that the LFQPOs and the HF QPOs occur on the same orbit.
	
	\item \textbf{same mechanism for 3:2 twin HF QPOs}\\
	The same mechanism is responsible for the 3:2 twin HF QPOs of all of the microquasars studies here, i.e. a single model can be applied to all of them.
\end{enumerate}

An upper bound on the specific angular momentum comes for the theory of the Kerr black hole -- it cannot be greater than one. The observation of a pair of simultaneous HF QPOs allows us to express the mass of the black hole as a function of the spin. With this metric and for the set of models considered here this function is monotonous.  It is increasing in the case of prograde motion of the hot spot which produces the HF QPOs. Hence, the upper bound on the spin gives a upper bound on the mass in the prograde case.

The presence of LF QPOs in the X-ray power density spectrum of a LMXB allows us to obtain a lower bound on its angular momentum. The values of the LF QPOs vary significantly with the evolution of the object. As we said above, according to the RP model LF QPOs result from the Lense-Thirring precession of matter inhomogeneities in the accretion disk, dubbed hot spots. The variance of the frequency values is produced by the variance of the orbital radii of the hot spots. The frequency of the LF QPOs is associated with the nodal precession frequency $\nu_{ \rm  nod}=|\nu_\phi-\nu_\theta|$. If the Kerr metric is assumed the nodal precession frequency is a function of three variables: $M$, $a$ and $r_{ \rm LF}$, where $r_{ \rm LF}$ is the radius of the orbit of the hot spot, i.e. the radius on which the LF QPOs originate. The last parameter can be obtained from the equation
\begin{eqnarray}
\nu_{ \rm  nod}(a,M,r_{ \rm LF})=\nu_{ \rm LF}^{ \rm  obs}, \label{eq_LF}
\end{eqnarray}
in which $\nu_{ \rm LF}^{ \rm obs}$ is an observed value of the LF QPOs. If the mass $M$ of the black hole is known, (\ref{eq_LF}) results in a $a-r_{ \rm LF}$ relation, the radius is an implicitly defined monotonously increasing function of the angular momentum. The radius of the orbit and, hence, the angular moment cannot be arbitrarily small. A lower bound comes from the natural requirement that the LF QPOs originate at an orbit whose radius $r_{ \rm LF}$ is greater than or at least equal to the radius of the innermost stable circular orbit $r_{ \rm ISCO}$. The radius $r_{ \rm ISCO}$ is function of $a$ alone, while $r_{ \rm LF}$ depends on both $a$ and $M$. In the limit $a\rightarrow0$, while $M$ is finite, the radius $r_{ \rm LF}(a,M)$ vanishes and is, hence, lower than $r_{ \rm ISCO}(a)$. For high enough values of the angular momentum $r_{ \rm LF}(a,M)\geq r_{ \rm ISCO}(a)$. A lower bound on the angular momentum $a_{ \rm LF}^{ \rm min}(M)$ is obtained from the equality $r_{ \rm LF}(a_{ \rm LF}^{ \rm min},M)= r_{ \rm ISCO}(a_{ \rm LF}^{ \rm min})$.  This equation is equivalent to the system composed of equation (\ref{eq_LF}) and the equation which defines the ISCO
\begin{eqnarray}
\nu_{ \rm  r}(a,r_{ \rm LF})=0 \label{eq_ISCO_def}.
\end{eqnarray}
Greater values of $\nu_{ \rm LF}^{ \rm  obs}$ yield greater lower bounds on the angular momentum. In other words, the greatest observed value of the LF QPOs in the spectrum of a given object gives the most stringent constraint on its $a_{ \rm LF}^{ \rm min}$. The function $a_{ \rm LF}^{ \rm min}(M)$ is monotonously increasing for prograde orbits.

In case $M$ is unknown $a_{ \rm LF}^{ \rm min}$ is also unknown. The minimum value of $a_{ \rm LF}^{ \rm min}$ which we denote as $a_{ \rm min}$, however, can be found if a pair of twin simultaneous HF QPOs has been observed in the X-ray power spectrum of a given BHB. With the observed HF QPOs one can compose the following algebraic system of equations
\begin{eqnarray}
&&\nu_{ \rm  L}(a,M,r_{ \rm HF})=\nu_{ \rm L}^{ \rm  obs}, \label{eq_L}\\
&&\nu_{ \rm U}(a,M,r_{ \rm HF})=\nu_{ \rm U}^{ \rm  obs} \label{eq_U}.
\end{eqnarray}
The explicit form of $\nu_{ \rm  L}(a,M,r_{ \rm HF})$ and $\nu_{ \rm  U}(a,M,r_{ \rm HF})$ depends on the choice of a model for the HF QPOs and of a metric.
If $M$ is known, the system above can be solved for the angular momentum of the BH $a$ and the radius of the orbit on which the HF QPOs occur  $r_{ \rm HF}$. For unknown values of $M$ the system (\ref{eq_L})-(\ref{eq_U}) results in a $a-M$ relation, i.e. the functions $a(M)$ and $r_{ \rm HF}(M)$ are defined implicitly. The presence of the $a-M$ degeneracy has been noticed in \cite{mass_angular_relation_Chi2_2010, mass_angular_relation_Chi2_2014} and \cite{moyataAN}.
For the radius and the angular momentum we have also the following constraints: $r_{ \rm HF}(M)\geq r_{ \rm ISCO}(a(M))$ and $a(M)\geq a_{ \rm LF}^{ \rm min}(M)$. The numerical analysis shows that in all cases studied below the former is satisfied. The latter is satisfied only for high enough masses. The equality gives us a lower bound on the mass $M_{ \rm min}$ and a lower bound on the angular momentum $a_{ \rm min}$: $a_{ \rm min}\equiv a(M_{ \rm min})= a_{ \rm LF}^{ \rm min}(M_{ \rm min})$.

To summarize what was said above, a lower bound on the angular momentum $a_{ \rm min}$ of a black hole whose X-ray power density spectrum contains the following triad of frequencies -- $\nu_{ \rm LF}^{ \rm  obs},\nu_{ \rm L}^{ \rm  obs}$ and $\nu_{ \rm U}^{ \rm  obs}$,  where the last two are simultaneously observed, is obtained from the requirement that $a(M_{ \rm min})= a_{ \rm LF}^{ \rm min}(M_{ \rm min})$. In general the unknown variables are five -- $a$, $a_{ \rm LF}^{ \rm min}$, $M$, $r_{ \rm LF}$, $r_{ \rm HF}$, while the equations are only four --  (\ref{eq_LF}), (\ref{eq_ISCO_def}), (\ref{eq_L}), (\ref{eq_U}). In the special case $M=M_{ \rm min}$, however, the relation $a=a_{ \rm LF}^{ \rm min}=a_{ \rm min}$ reduces the number of the unknown variables and it turns out that we can find $a_{ \rm min}$ with the following recipe. The system
\begin{eqnarray}
&&{\bar{\nu}_{ \rm L}(a_{ \rm min},r_{ \rm HF})\over \bar{\nu}_{ \rm U}(a_{ \rm min},r_{ \rm HF})}={\nu_{ \rm U}^{ \rm  obs} \over \nu_{ \rm L}^{ \rm  obs}}, \label{eq_UL}\\
&&{\bar{\nu}_{ \rm  L}(a_{ \rm min},r_{ \rm HF})\over \bar{\nu}_{ \rm LF}(a_{ \rm min},r_{ \rm LF})}={\nu_{ \rm L}^{ \rm  obs} \over \nu_{ \rm LF}^{ \rm  obs}}, \label{eq_LLF}\\
&&\bar{\nu}_{ \rm  r}(a_{ \rm min},r_{ \rm LF})=0 \label{eq_ISCO}.
\end{eqnarray}
is independent of $M$. The bar designates the part of the expression for the frequencies that is independent of $M$
\begin{eqnarray}
\nu_i(a,M,r)={\bar{\nu}_i(a,r)\over M}, \quad\quad i=U,\,L,\,LF,\,r.
\end{eqnarray}
The system (\ref{eq_UL})--(\ref{eq_ISCO}) has three equations and three unknowns: $a_{ \rm min}, \, r_{ \rm HF}$ and $r_{ \rm LF}$. It is solved numerically. Once $a_{ \rm min}$ and $r_{ \rm HF}$ are found $M_{ \rm min}$ can be obtained from $\nu_{ \rm  L}(a_{ \rm min},M_{ \rm min},r_{ \rm HF})=\nu_{ \rm L}^{ \rm  obs}$ (or $\nu_{ \rm U}(a_{ \rm min},M_{ \rm min},r_{ \rm HF})=\nu_{ \rm U}^{ \rm  obs}$).

The upper bound for the mass comes from the system of equations (\ref{eq_L}) and (\ref{eq_U}) in which the substitution $a=1$ is made.

Bellow we designate the lower bound on the mass $M_{ \rm min}$ by $M_l$, or $M_{ \rm min} \equiv M_l$ .

\section{Mass Bounds for the Studied Objects}
\begin{figure}
	\centering
	\includegraphics[ width=0.6\textwidth,keepaspectratio]{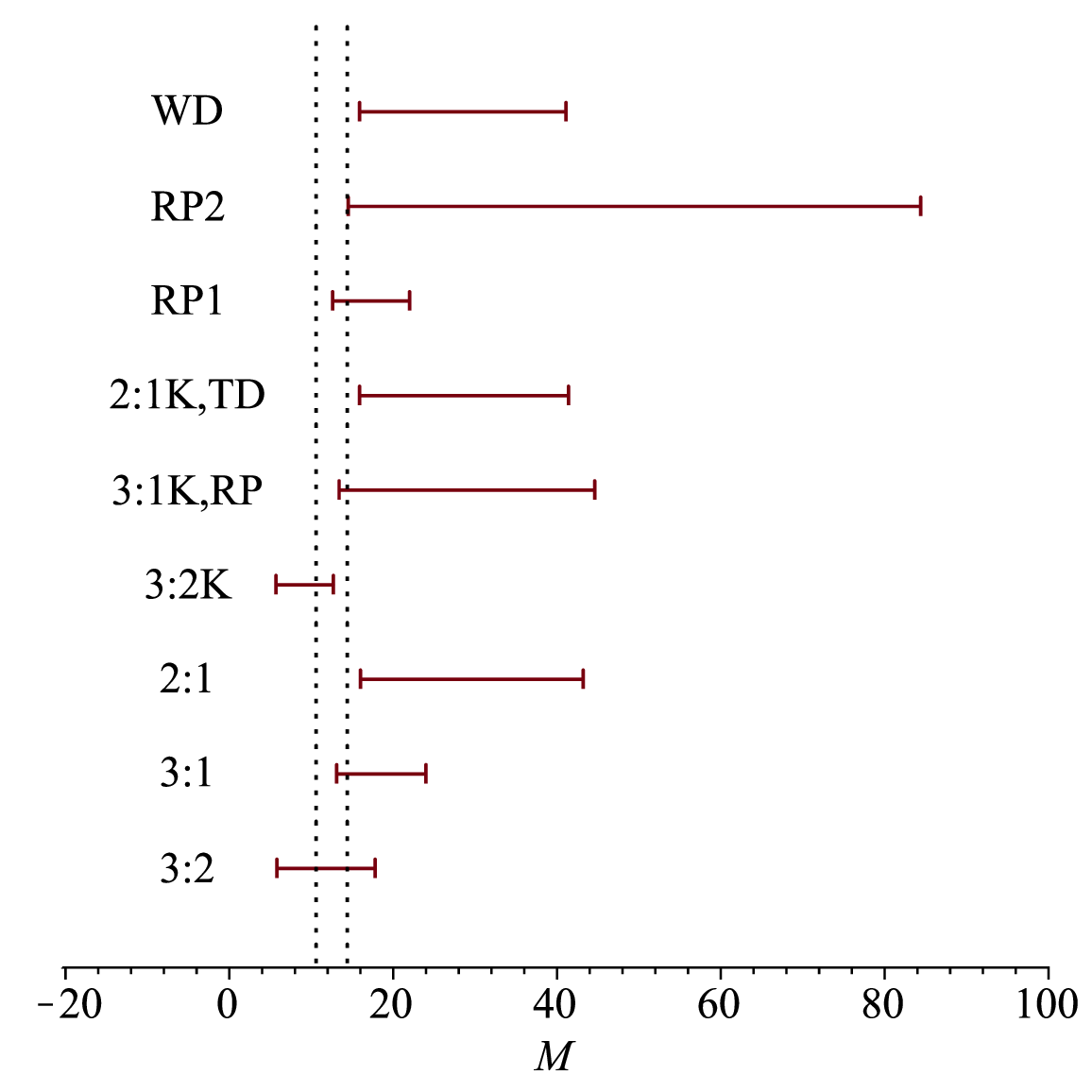}
	\caption{Mass bounds for GRS 1915+105. The models are designated on the left. The dotted lines represent the reference masses.}\label{GRS}
\end{figure}

\begin{figure}[t]
	\centering
	\includegraphics[ width=0.6\textwidth,keepaspectratio]{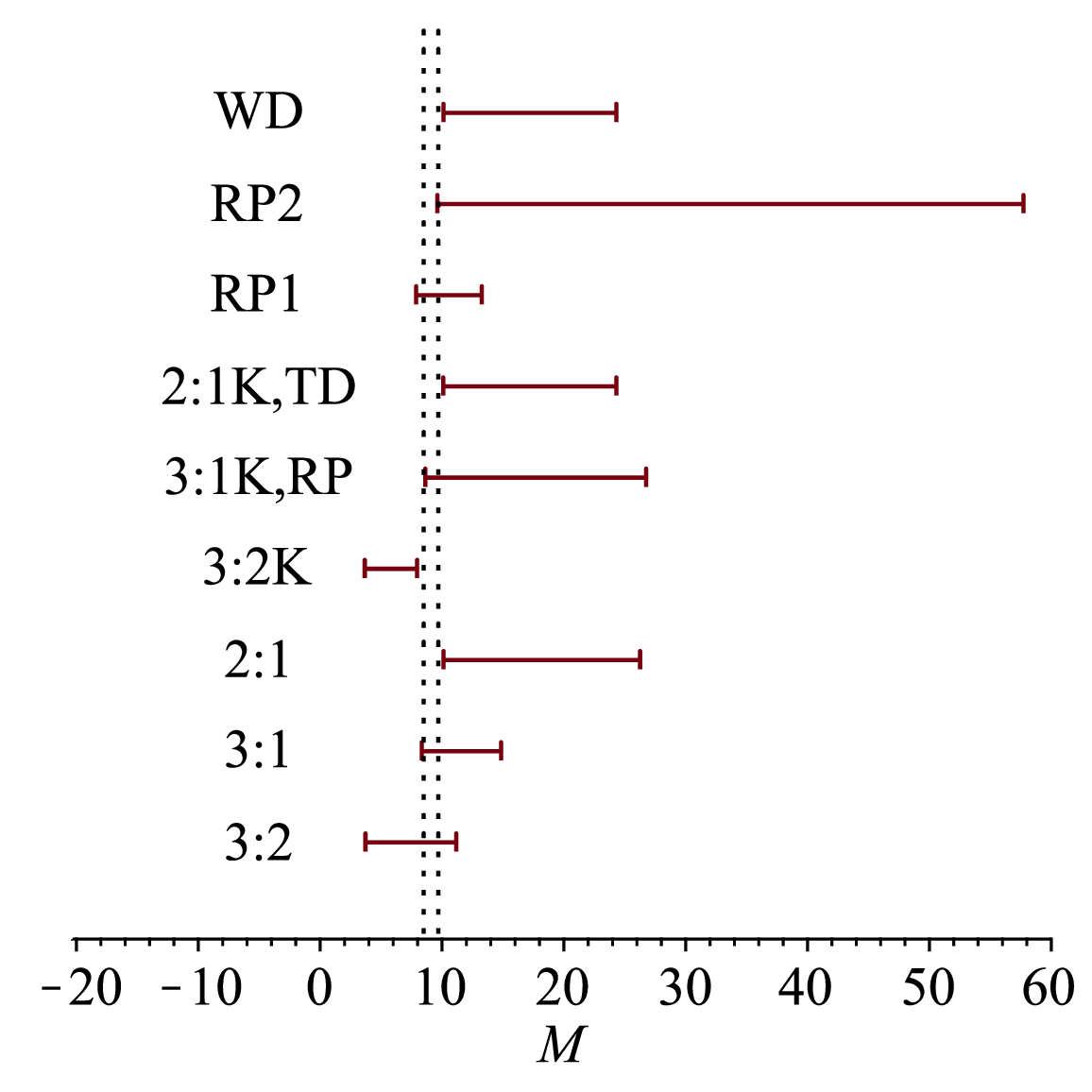}
	\caption{Mass bounds for XTE 1550-56. The models are designated on the left. The dotted lines represent the reference masses.}\label{XTE18}
\end{figure}

\begin{figure}
	\centering
	\includegraphics[ width=0.6\textwidth,keepaspectratio]{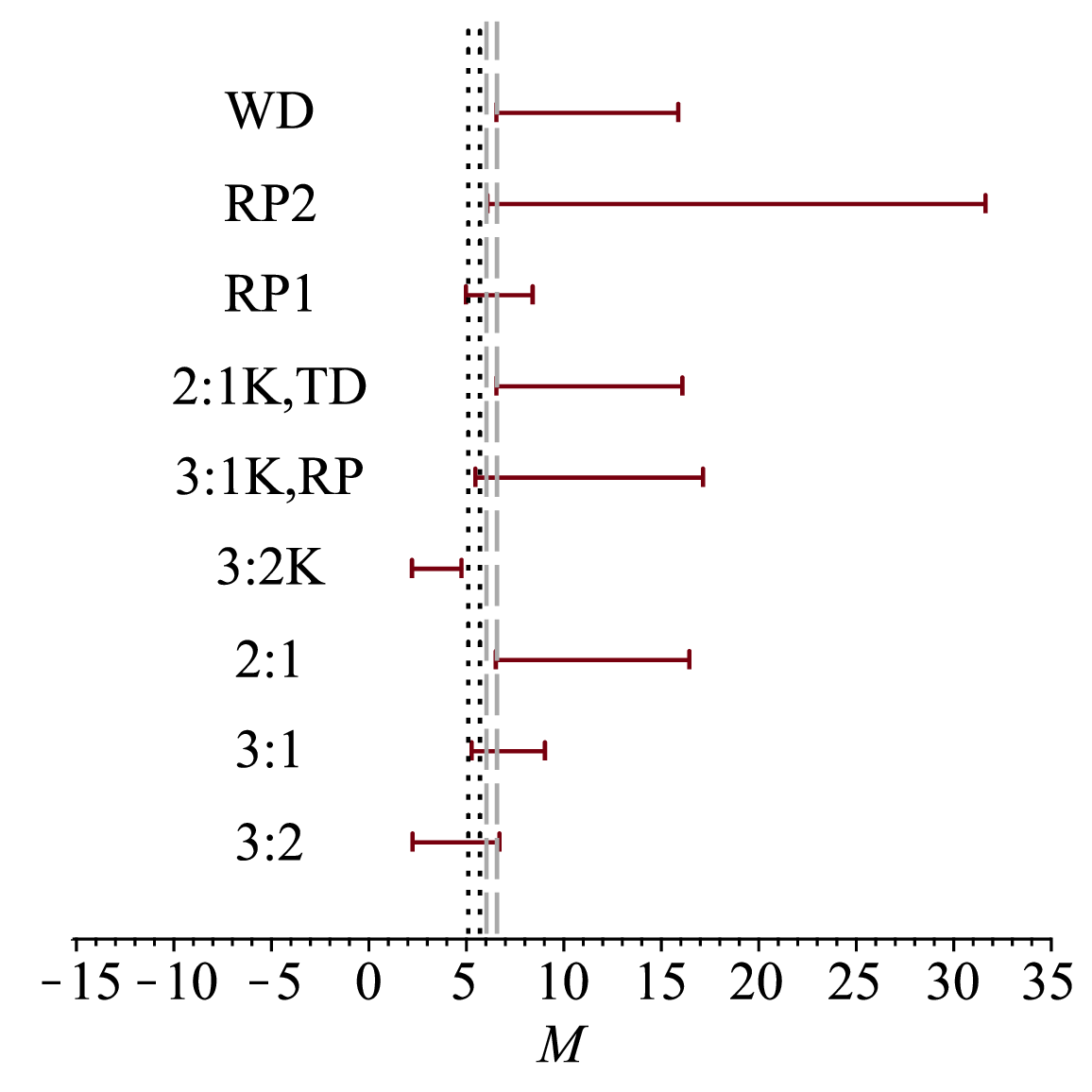}
	\caption{Mass bounds for GRO 1655-40. The models are designated on the left. The dotted lines represent the reference masses. The alternative reference mass interval is given by the dashed gray line.}\label{GRO}
\end{figure}

The mass bounds that the different models impose on the studied black holes are given in the current section. They have been obtained with the application of the method described in Section \ref{Method}. The predicted intervals and the reference masses are presented on Figures \ref{GRS}, \ref{XTE18} and \ref{GRO} for GRS  1915+105, XTE 1550-564 and  GRO~1655-40, respectively. On these figures, the vertical dotted lines represent the reference masses. The numerical values for the masses and spins are given in Table~\ref{values}.
\begin{table}
	\centering
	\small
	\caption{Bounds on the mass and spin}
	\begin{tabular}{|l|c|c|c|c|c|c|}
		\cline{2-7}
		\multicolumn{1}{c|}{}&\multicolumn{2}{|c|}{GRS 1915+105}&\multicolumn{2}{|c|}{XTE 1550-564}&\multicolumn{2}{|c|}{GRO 1655-40}\\
		\hline
		Model & $a_{ \rm min}$  &$M/M_\odot$ & $a_{ \rm min}$  &$M/M_\odot$& $a_{ \rm min}$  &$M/M_\odot$\\
		\hline
		3:2            & 0.13 & 5.8  --  17.9&0.18&3.7 -- 11.2&0.17&2.2 -- 6.7\\
		3:1            & 0.25 & 13.1  --  24.0&0.31&8.3 -- 14.9 &0.31&5.3 -- 9.0\\
		2:1            & 0.28 & 15.9  --  43.2 &0.35&10.1 -- 26.3&0.36&6.5 -- 16.4\\
		3:2~K          & 0.13 & 5.7  --  12.7&0.17&3.7 -- 8.0&0.17&2.2 -- 4.8\\
		RP, 3:1~K      & 0.25 & 13.4  --  44.6&0.32&8.6 -- 26.8&0.32&5.5 -- 17.1\\
		TD, 2:1~K      & 0.28 & 15.9  --  41.4 &0.35&10.1 -- 24.3&0.36&6.5 -- 16.1\\
		RP1            & 0.24 & 12.6  --  22.0&0.30&7.9 -- 13.3 &0.30&5.0 -- 8.4\\
		RP2	           & 0.27 & 14.5  --  84.4 &0.34&9.6 -- 52.7&0.34&6.1 -- 31.6\\
		WD	           & 0.28 & 15.9  --  41.1&0.35&10.1 -- 24.3&0.36&6.5 -- 15.9\\
	\hline
	\end{tabular}\label{values}
\end{table}

Some general properties of the models are seen immediately. The model RP2 allows the biggest mass tolerance, while the 3:2~K model is the most restrictive.

\subsection{GRS  1915+105}

The masses predicted by the different models for this black hole are in the range $5.8\leq M/M_\odot\leq 84.4$.  The 3:2~K model predicts the lowest value $5.8 M_\odot$, while the RP2 model gives the maximum value for the mass $84.4M_\odot$. One of the models distinctly satisfy the predicted in the literature diapason -- 3:2. The following models -- 3:1; 3:2~K; RP, 3:1~K and RP1 cover this range only partially. The model 3:2~K  provides the shortest value range,  $5.7\leq M/M_\odot\leq12.7$. 
\subsection{XTE 1550-564}
The masses predicted by the different models are in the range $3.7\leq M/M_\odot\leq 52.7$  The model 3:2 gives the lowest value of the mass $3.7M_\odot$, while the model RP2 gives the maximum value $52.7 M_\odot$. Four of the nine models entirely cover the range given in Table~\ref{objects} but predict larger diapason of masses -- 3:2; 3:1; 3:1~K; RP and RP1. 3:2~K again has the shortest range of values $3.7\leq M/M_\odot\leq8.0$.
The models 2:1; 3:2~K; TD, 2:1~K; RP2 and WD,  do not cover the reference range even partly. 

\subsection{GRO 1655-40}
The masses predicted by the different models are in the range $2.2\leq M/M_\odot\leq 31.6$.  The 3:2 model and the 3:2~K model predict the lowest value for the mass $2.2M_\odot$. The model RP2 gives the maximum value for the mass $31.6 M_\odot$. There are two intervals of mass in the literature, $5.1\leq M/M_\odot\leq5.7$  and $6.03 \leq M/M_\odot\leq 6.57$. Four of the nine models provide a mass interval overlapping partially with both ranges obtained by other authors  -- 3:2; 3:1;  RP, 3:1~K and RP1. From them  RP1 shows the shortest value range --  $5.0\leq M/M_\odot\leq8.4$. RP2 satisfies only the second reference interval. The 2:1 and 3:2~K models give no solution in the reference range. The models TD, 2:1~K and WD predict identical mass intervals which, however, do not match those in the literature. 

\section{Quality of the models}
In the current section
we propose a simple method to estimate the quality of the models. We would like to know what is the average success of the models. What do we mean by ``average success''? Each of the models provides an interval of values for the masses of the observed objects. It appears that the prediction of a given model and the mass obtained through dynamical measurement are in agreement in some cases and in conflict in others. One might ask, which situation is more likely to occur? Is a given model more likely to provide correct mass bounds or not? Can failures be attributed to random errors or are they systematic and should be treated as an indication of a bad model? In the current section we propose a simple statistical approach to this problem which allows us to identify the models for which the discrepancies between predictions and observations are either too big or too frequent.

In a typical situation we would have a sample of experimental values for the mass of a given object. The sample mean is taken as the best estimate of the mass. Here, we are confronted with a different situation. We have one measurement for each of the three objects. Their average does not give us any useful information. We suggest, however, that there is something that the three objects, and all other objects which display HF QPOs in 3:2 ratio and LF QPOs in their X-ray spectra, might have in common and that can be averaged -- the relative position of their reference (or observed) mass  in the interval of masses that a given model for the HF QPOs provides for them.

Let us explore this idea in more details. We make the hypothesis that the models which we want to test should show some tendency. The reference masses of all three of the objects are to the left of the center of the interval coming from the RP model, for example. Our aim is to find the tendencies that the different models exhibit. In other words we would like to know whether a given model is more likely to underestimate or overestimate the masses of the objects. In which part of the interval do  the observations occur most frequently? Is there such point in the interval that ``attracts'' the observations? These questions can be expressed in a more formal way. Are all masses in the interval provided by a given model equally probable? What is the probability distribution of the values in the interval? Which of the values in the interval gives the best estimate of the mass? Since the probability distribution of the masses is not known we resort to statistics and averaging.

\subsection{Rescaled masses}

\begin{figure}
	\centering
	\includegraphics[width=0.6\textwidth,keepaspectratio]{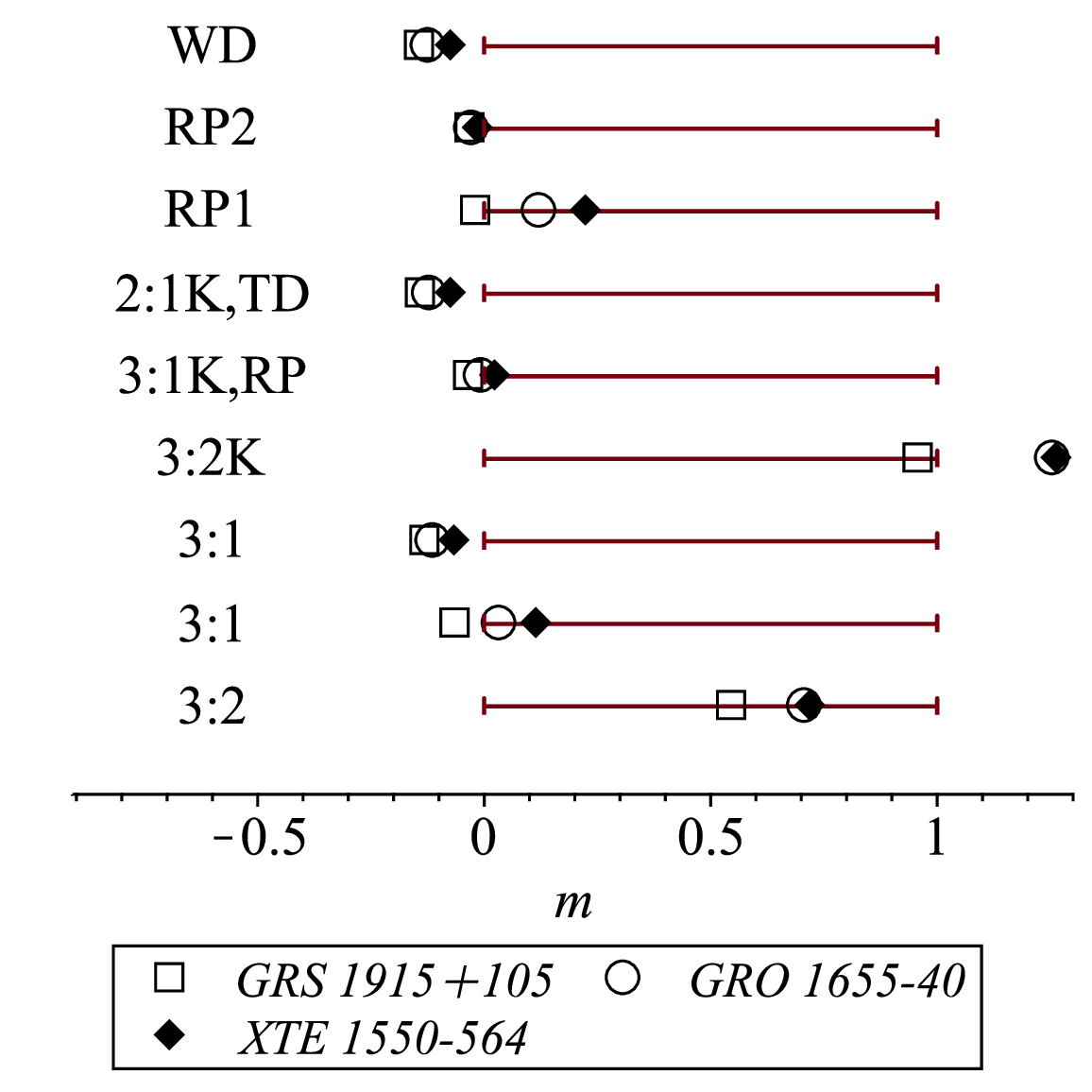}
	\caption{The unit intervals and the relative positions of the observed masses of the objects for all of the studied models. Each model is represented by a different symbol.}\label{relative_positions}
\end{figure}

In order to allow averaging we project the mass intervals that a given model provides for the three objects on the unit interval through the following translation
\begin{eqnarray}\label{projection}
m={M-M_l \over \Delta M}, 
\end{eqnarray}
where $M_l$ and $M_u$ are, respectively, the lower and the upper mass bounds and $ \Delta M=M_u-M_l$. For $M\in[M_l, M_u]$ the new variable $m$ takes values in the interval $[0, 1]$. The relative position, or coordinates, of the observed mass is defined as $m_{\rm obs}=m(M^{\rm obs})$. It is in the interval $[0, 1]$ only when  $M^{\rm obs}\in[M_l, M_u]$. The cases $m_{\rm obs}<0$  and $m_{\rm obs}>1$ indicate a conflict between the prediction of the given model and the observed mass of the object. The unit intervals and the relative positions of the observed masses for all of the studied models are presented in Figure \ref{relative_positions}.
The successful hits are easily recognizable.
\subsection{Sample mean and sample variance}
Let us suppose that we could add more and more objects that share common properties to our sample. Then the values of $m_{\rm obs}$ would span an interval which would coincide with the unit interval in the ideal case\footnote{$m_{\rm obs}$ would cover the entire unit interval, if the spin of the black hole which has produced a give set of values for the HF and LF QPOs could take any value in the interval $[a_{\rm min}$, 1].}. Since no model is perfect the two intervals would not overlap completely. Significant discrepancy between the prediction and the observations would be an indication of a bad model. Let us assume that the expectation value of $m_{\rm obs}$, i.e. the population mean $\mu_{m_{\rm obs}}$, was known. It represents the average prediction of a given model. The inverted  transformation  (\ref{projection}) allows us to find the best estimate of the mass: $M_{\rm best\, estimate}=M(\mu_{m_{\rm obs}})$. If $\mu_{m_{\rm obs}}<0$ or $\mu_{m_{\rm obs}}>1$ then the model tends to, respectively, overestimate or underestimate the masses of the objects. In case $\mu_{m_{\rm obs}}=0.5$ the model is in complete agreement with the observations. Since the population is unknown we will work with the available sample which consists of only three objects.

The average relative position $\bar{m}_{\rm obs}$\footnote{See appendix \ref{app_stat} for this and all the other formulas used in the current section.}, i.e. the sample mean,  serves as an estimate of the population mean $\mu_{m_{\rm obs}}$,  $\widehat{\mu}_{m_{\rm obs}}=\bar{m}_{\rm obs}$ as usual. The sample's standard deviation SD  shows how strong the tendency that the models reveal is. Small values of SD mean that all the observations group in a narrow interval. The standard error of the sample mean (SEM), which tells us how far is the sample mean likely to be from the population mean, is given by $SD/\sqrt{N}$, where $N=3$ in our case.

The results from the averaging of these quantities over the available sample of three objects for each of the nine studied models is given in Table~\ref{stats}.
\begin{table}[t]
	\centering
	\caption{Statistics}
	\begin{tabular}{|c|c|c|c|}
		\hline
		Model & $\bar{m}_{\rm obs}\pm {\rm SEM}$ & SD&$\sigma_{\bar{m}_{\rm obs}}$\\
		\hline
		3:2&$0.66\pm0.06$&0.1&0.09\\
		3:1&$0.03\pm0.05$&0.09&0.10\\
		\textbf{2:1}&$-\textbf{0.10}\pm\textbf{0.02}$&0.03&0.04\\
		\textbf{3:2~K}&$\textbf{1.16}\pm\textbf{0.10}$&0.17&0.16\\
		RP, 3:1~K&$0.00\pm0.02$&0.03&0.04\\
		\textbf{TD, 2:1~K}&$-\textbf{0.11}\pm\textbf{0.02}$&0.03&0.04\\
		RP1&$0.11\pm0.07$&0.12&0.12\\
		\textbf{RP2}&$-\textbf{0.02}\pm\textbf{0.01}$&0.01&0.02\\
		\textbf{WD}&$-\textbf{0.11}\pm\textbf{0.02}$&0.04&0.04\\
		\hline
	\end{tabular}\label{stats}
\end{table}
The models are listed in the first column. Column two represents the sample mean and its standard error, and the standard deviation of the experimental points is in the third column. The fourth column is commented bellow.
\subsection{Viability of the models}
The models for which the sample mean is outside of the unit interval even when the SEM is taken into account, i.e. they are in clear conflict with the observations, are marked in bold. The third column of Table~\ref{stats} reveals that the tendencies are rather strong since the scatter of the points is only $17\%$ at most (See the data for model 3:2~K.). As it can be seen, our results disfavor five of the nine groups of models: 2:1; 3:2~K; TD, 2:1~K; RP2; WD. The viable models in decreasing order of their quality\footnote{Here, by quality we mean the distance of the average position from the center of the interval -- the smaller the distance, the greater the quality. Once $\mu_{m_{\rm obs}}$ corresponding to a given model is known we can apply it to obtain the best estimate of the mass, even if the model demonstrates poor quality. We would know what correction should be made in the prediction coming form the model in question. Loosely speaking, $\left|\mu_{m_{\rm obs}}-1/2\right|$ is the bias of the models.} are: 3:2; RP1; 3:1 and RP, 3:1~K.
\subsection{The effect of uncertainties in the reference masses}\label{mass_uncertainty}
The estimates for the masses of the objects that we use as reference might vary significantly from one study to the other. We refer the reader to \cite{GRS_12_4_mass} and \cite{GRS_the_not_so_massive} for different estimates of the mass of GRS 1915+105, and to \cite{GRO_Beer_54}  and \cite{GRO_6_3_mass} for different estimates of the mass of GRO 1655-40, for example.

With this in mind, we evaluate the effect that uncertainties in the reference masses $\sigma_{M^{\rm obs}}$ have on the results given in the second column of Table~\ref{stats}, which are the basis for our conclusions about the viability of the models. Variance within one Solar mass, i.e. $\sigma_{M^{\rm obs}}=1 M_\odot$, for all of the three microquasars is reasonable. The results for the uncertainties of the average positions of the observed masses $\sigma_{\bar{m}_{\rm obs}}$ are given in the last column of Table~\ref{stats}.

Combined unfavorable contribution of SEM and $\sigma_{\bar{m}_{\rm obs}}$ would have no effect on the conclusion concerning the viability of the following models: 3:2; 2:1; TD, 2:1~K; WD. The possibility that errors could change the conclusions about the other five groups of model -- 3:1; 3:2~K; RP, 3:1~K; RP1; RP2 -- cannot be discarded.

\section{Conclusion}


Two of the models, 3:2 and 3:2~K, show systematic tendency to underestimate the masses of the objects. The reference mass is to the right of the center of the mass intervals obtained with them for all three of the objects. 
On the contrary, the rest of the models -- 3:1; 2:1; RP, 3:1~K; TD, 2:1~K; RP1; RP2 and WD,   seem to overestimate the masses of the studied objects.
Relatively moderate mass bounds, neither too high nor too low, are provided by the following set of models -- 3:2; 3:1; RP1.
These models appear to be favored by the current study.

Of all the models the 3:2~K and the RP1 are the most restrictive and still the predictions of the latter are rather acceptable\footnote{The more restrictive a given model is, the more likely a conflict with the referential interval is to occur.}. The RP2 model is the least restrictive. 

With a few exceptions, the lower bound on the specific angular momentum of the central object that a given model imposes, $a_{ \rm min}$, varies slightly from one object to the other.

The statistical approach for the assessment of the average success of the models reveals that conflicts the mass bounds provided by the following models: 2:1; 3:2 K; TD, 2:1~K; RP2 and WD, on the one side, and the reference masses, on the other side, occur more often than not. The viable models, in increasing order of their quality, are: 3:2; RP1; 3:1 and RP, 3:1~K.

The conclusions based on the averaging of the observations for the three microquasars are sensitive to our knowledge of the precise values of the reference masses (denoted here as $M^{\rm obs}$). For uncertainties lower than or at most equal to one Solar mass our conclusions for  3:2; 2:1; TD, 2:1~K; WD would not be affected.

\appendix
\section{Epicyclic frequencies}\label{app_freqs}
The explicit form of the orbital frequency $\nu_{\rm \phi}$ and the two epicyclic frequencies -- the radial $\nu_r$ and the vertical $\nu_{\theta}$ -- for the Kerr black hole have been obtained for the first time in \cite{AlievGaltsov1, AlievGaltsov2} but can be found also, for example, in \cite{AlievKerr}
\begin{eqnarray}
&&\nu_{\rm \phi} =\left({c^3\over 2\pi GM}\right)\frac{ 1}{ r^{3/2} + a}, \\
&&\nu_{r}^2 = \nu_{\rm \phi}^2\, \left( 1-\frac{6 }{r} -\frac{3
	a^2}{r^2} + {8 a\over r^{3/2}}\right),\\
&&\nu_{\theta}^2= \nu_{\rm \phi}^2\, \left(1
+\frac{3 a^2}{r^2} - {4 a \over r^{3/2}} \right).
\end{eqnarray}
A change in the orientation of the orbit (direction of rotation of the hot spot) is equivalent to a change of the direction of rotation of the central object, i.e. a change in the sign of $a$. In this paper all the masses are scaled with the Solar mass, the radii are scaled with the gravitational radius $r_{\rm g}\equiv G M/c^2$,  and the specific angular momentum $a\equiv J/c M^2$ is used.
\section{Statistical formulas}\label{app_stat}
The formulas for the sample mean $\bar{m}_{\rm obs}$ and the sample standard deviation SD are:
\begin{eqnarray}
&&\bar{m}_{\rm obs}={1\over N}\sum_{i=1}^N m_{{\rm obs},i},\\
&&SD=\sqrt{{1\over N-1}\sum_{i=1}^N \left(\bar{m}_{\rm obs}-m_{{\rm obs},i}\right)^2}\,\,.
\end{eqnarray}
In order to obtain the uncertainty of the average position of the observed masses $\sigma_{\bar{m}_{\rm obs}}$ which is due to the uncertainties of the observed (reference) masses $\sigma_{M^{\rm obs}_i}$ we apply a standard uncertainty propagation formula:
\begin{eqnarray}
&&\sigma_{\bar{m}_{\rm obs}}^2=\sum_{i=1}^N \left({\partial \bar{m}_{\rm obs}\over \partial M^{\rm obs}_i}\right)^2 \sigma_{M^{\rm obs}_i}^2=
{1\over N^2}\sum_{i=1}^N {1\over \Delta M_i^{\,2}}\, \sigma_{M^{\rm obs}_i}^2.
\end{eqnarray}
Here the index $i$ enumerates the three microquasars: GRS 1915+105, GRO 1655-40 and  XTE 1550-564, and $N=3$.
\nocite{*}

\begin{acknowledgement}
I.S. would like to thank his wife for the support, Dr. Sava Donkov for reading the manuscript and for the numerous discussions on the subject, prof. Stoytcho Yazadjiev for drawing his attention to the subject of quasiperiodic oscillations of black holes and neutron stars. The research is partially supported by the the Bulgarian National Science Fund under Grant No N 12/11 from 20 December 2017.
\end{acknowledgement}

\end{document}